\documentclass[preprint]{aastex}
\usepackage{amsmath}
\usepackage{xspace}
\usepackage{bm}

\newcommand{\hrms}{\ensuremath{{h_\mathrm{rms}}}\xspace}
\newcommand{\Msun}{\ensuremath{{M_\odot}}\xspace}
\newcommand{\Rsun}{\ensuremath{{R_\odot}}\xspace}
\newcommand{\dsun}{\ensuremath{{d_\odot}}\xspace}
\newcommand{\Ylm}{\ensuremath{{Y^{\ell m}}}\xspace}
\newcommand{\Ilm}{{\ensuremath{{\mathcal{I}^{\ell m}}}}\xspace}
\newcommand{\Slm}{\ensuremath{{\mathcal{S}^{\ell m}}}\xspace}
\newcommand{\Itwom}{\ensuremath{{\mathcal{I}^{2 m}}}\xspace}
\newcommand{\Stwom}{\ensuremath{{\mathcal{S}^{2 m}}}\xspace}
\newcommand{\etaks}{\ensuremath{{\eta(k_s)}}\xspace}
\newcommand{\Ogw}{\ensuremath{{\Omega_\mathrm{gw}}}\xspace}
\newcommand{\Ogwsquared}{\ensuremath{{\Omega_\mathrm{gw}^2}}\xspace}
\newcommand{\Tobs}{\ensuremath{{T_\mathrm{obs}}}\xspace}
\newcommand{\xx}{\ensuremath{{\bm{x}}\xspace}}
\newcommand{\vv}{\ensuremath{{\bm{v}}\xspace}}
\newcommand{\kk}{\ensuremath{{\bm{k}}\xspace}}
\newcommand{\nn}{\ensuremath{{\bm{n}}\xspace}}
\newcommand{\mm}{\ensuremath{{\bm{m}}\xspace}}

\begin{document}

\title{Stochastic microhertz gravitational radiation from stellar convection}
\author{M. F. Bennett and A. Melatos}
\affil{School of Physics, University of Melbourne, Parkville, VIC 3010, Australia}
\email{mfb@unimelb.edu.au}

\begin{abstract}
    High-Reynolds-number turbulence driven by stellar convection in main-sequence stars generates stochastic gravitational radiation.  We calculate the wave-strain power spectral density as a function of the zero-age main-sequence mass for an individual star and for an isotropic, universal stellar population described by the Salpeter initial mass function and redshift-dependent Hopkins-Beacom star formation rate.  The spectrum is a broken power law, which peaks near the turnover frequency of the largest turbulent eddies.  The signal from the Sun dominates the universal background.  For the Sun, the far-zone power spectral density peaks at $S(f_\mathrm{peak}) = 5.2 \times 10^{-52}~\mathrm{Hz}^{-1}$ at frequency $f_\mathrm{peak} = 2.3 \times 10^{-7}~\mathrm{Hz}$.  However, at low observing frequencies $f < 3 \times 10^{-4}~\mathrm{Hz}$, the Earth lies inside the Sun's near zone and the signal is amplified to $S_\mathrm{near}(f_\mathrm{peak}) = 4.1 \times 10^{-27}~\mathrm{Hz}^{-1}$ because the wave strain scales more steeply with distance ($\propto d^{-5}$)  in the near zone than in the far zone ($\propto d^{-1}$).  Hence the Solar signal may prove relevant for pulsar timing arrays.  Other individual sources and the universal background fall well below the projected sensitivities of the Laser Interferometer Space Antenna and next-generation pulsar timing arrays.  Stellar convection sets a fundamental noise floor for more sensitive stochastic gravitational-wave experiments in the more distant future.
\end{abstract}

\keywords{gravitational waves --- turbulence --- convection --- stars: general --- stars: interior --- Sun: interior}

\section{Introduction}\label{sec:intro}

Stochastic gravitational-wave backgrounds arise from the superposition of many unresolved point sources, e.g., compact object binaries \citep{farmer03, sesana08, rosado11}, supernovae \citep{ferrari99a, coward01}, magnetars \citep{regimbau06, marassi11}, rotating neutron stars \citep{ferrari99b, regimbau01}, and pulsar glitches \citep{warszawski12}.  Point source backgrounds establish a noise floor for detection of extended backgrounds generated by fundamental processes early in the life of the Universe, e.g., cosmic strings \citep{damour05, siemens07}, inflation \citep{starobinskii79, bar-kana94}, or primordial turbulence \citep{kosowsky02, gogoberidze07}.  Recently, a search in two years of data from the fifth science run (S5) of the Laser Interferometer Gravitational-wave Observatory (LIGO) placed an upper limit on the gravitational-wave energy density in the Universe at 100 Hz, which supplanted previous limits from Big Bang nucleosynthesis and the cosmic microwave background \citep{abbott09}.

Relative to some of the above sources, main-sequence stars and their interiors are well understood.  In particular, the Sun has been studied extensively through helioseismology \citep[e.g.,][]{christensen-dalsgaard02}.  In this paper, we calculate the stochastic gravitational radiation emitted by convection in main-sequence stars, taken individually and collectively.  High-Reynolds-number turbulence is instantaneously nonaxisymmetric and therefore generates gravitational radiation even though it is axisymmetric when averaged over many turnover times \citep{melatos10, lasky13}.  We pay particular attention to the Sun, where convection can be observed indirectly through helioseismology and directly by Doppler imaging of granulation at the Solar surface \citep{miesch05}.  Previous studies \citep{cutler96, polnarev09} calculated the space-time perturbations generated by normal oscillation modes of the Sun and found that low-order modes, whose energy exceeds $\sim 10^{30}$ erg, may be detectable with the Laser Interferometer Space Antenna (LISA).

The paper is structured as follows.  In Section \ref{sec:stochastic_gw_signal}, we derive analytically the power spectral density of the quadrupole radiation emitted by a convective main-sequence star as a function of its mass.  The spectrum is evaluated for a selection of representative objects in Section \ref{sec:individual_sources} including the near-zone effects for the Sun.  In Section \ref{sec:stochastic_background}, we calculate the stochastic gravitational-wave background from an isotropic distribution of stars throughout the Universe and compare the predicted signal to the LISA noise curve.  The paper concludes by discussing critically the assumptions behind our idealized model in Section \ref{sec:conclusion}.

\section{Stochastic gravitational-wave signal}\label{sec:stochastic_gw_signal}

\subsection{Power spectral density}\label{sec:power_spectral_density}

In the transverse-traceless gauge, the gravitational-wave strain at a distance $d$ from a source is given by
\begin{equation}\label{eq:hTT}
    h_{jk}^\mathrm{TT} = \frac{G}{c^5 d}\sum_{\ell = 2}^{\infty} \sum_{m = -\ell}^{\ell} \frac{\partial^\ell \Slm(t)}{\partial t^\ell} T^{B2,\ell m}_{jk}~,
\end{equation}
where \Slm is the current multipole of order $(\ell, m)$ written as a function of the retarded time $t$, and $T^{B2,\ell m}_{jk}$ is a tensor spherical harmonic which describes the angular dependence of the radiation field and is itself transverse-traceless \citep{thorne80}.  \citet{melatos10} evaluated equation \eqref{eq:hTT} for shear-driven turbulence in a differentially rotating star, where it is permissible to neglect the mass multipoles \Ilm in favor of the current multipoles \Slm.  In main-sequence stars, where the convection speed and the adiabatic sound speed are smaller, the mass multipoles can also become important.  We discuss this point and estimate a correction factor in Section \ref{sec:mass_quadrupole}.

The two-time autocorrelation function $C(\tau)$ reflects the statistical properties of the turbulence and is related to the strain through
\begin{equation}\label{eq:autocorrelation_to_wave_strain}
    C(\tau) = \left\langle h_{jk}^\mathrm{TT}(t) h_{jk}^\mathrm{TT}(t')^* \right\rangle~,
\end{equation}
with $\tau = t - t'$, where $\langle \ldots \rangle$ represents the ensemble average over realizations of the turbulence.  We assume that the turbulence is isotropic and stationary, with the standard Kraichnan form for the velocity correlation function [equation (2) in \citet{melatos10}] and a Kolmogorov spectrum with energy per unit wavenumber $E(k) = k^2 P(k) \propto k^{-5/3}$ [equation (5) in \citet{melatos10}].  Simulations of three-dimensional turbulent convection produce results consistent with a Kolmogorov power law \citep{chan96, porter00, arnett09}.  Under these assumptions, equation \eqref{eq:autocorrelation_to_wave_strain} reduces to \citep{melatos10}
\begin{eqnarray}\label{eq:autocorrelation}
    \nonumber \frac{C(\tau)}{\hrms^2} &=& \left[ 1 - \frac{7 \pi \etaks^2 \tau^2}{2} \right] \exp \left[-\frac{\pi\etaks^2 \tau^2}{4} \right] \\
    && \qquad + 2 \pi^2 \etaks^3 \tau^3 \left\{ \mathrm{Erf}\left[ \frac{\pi^{1/2} \etaks \tau}{2} \left(\frac{k_d}{k_s}\right)^{2/3} \right] - \mathrm{Erf} \left[ \frac{\pi^{1/2} \etaks \tau}{2} \right] \right\}~,
\end{eqnarray}
where $k_s$ and $k_d$ are the stirring and viscous dissipation wavenumbers respectively, between which the Kolmogorov power law extends, $\epsilon$ is the power injected per unit enthalpy, and
\begin{equation}\label{eq:eddy_turnover}
    \eta(k) = \ (2 \pi)^{-1/2} \epsilon^{1/3} k^{2/3}
\end{equation}
is the reciprocal of the eddy turnover time at wavenumber $k$.  

The mean-squared wave strain evaluates to $C(0) = 0.59 G^2 \rho^2 R^8 \epsilon^2 / (c^{10} d^2)$ for a star with uniform density $\rho$ and radius $R$, if the whole interior is turbulent, and the size of the largest eddies is $R$, as in a differentially rotating neutron star \citep{melatos10}.  For main-sequence convection, $C(0)$ scales slightly differently.  From $\Stwom \propto \int d^3\xx \, r^2 \xx \cdot \mathrm{curl} (\rho \vv)$, where $\vv(\xx, t)$ is the velocity field in the star, we obtain
$[C(0)]^{1/2} \propto d^2 \Stwom / dt^2 \propto \rho V_\mathrm{turb} \langle r^3 \rangle l^{-1} v(l) \eta(2\pi/l)^2$, where $V_\mathrm{turb}$ is the turbulent volume, $v(l)$ is the typical turbulent speed in an eddy of linear dimension $l$, $\langle r^3 \rangle$ is the mass-weighted, mean-cube radius, and the curl operator is replaced approximately by $l^{-1}$ and $d/d\tau$ by $\eta(2\pi/l)$.  Stellar convection occurs either in the core or in an outer shell, depending on the zero-age main-sequence mass.  In general, the radial depth of the convective region $\Delta R$ is a function of zero-age main-sequence mass $M$ [e.g., see Figure 22.7 in \citet{kippenhahn90}].  For simplicity, we assume $\Delta R = R / 2$, and evaluate the mass-weighted, mean-cube radius $\langle r^3 \rangle$ in the core ($0 \le r \le R/2$) and outer shell ($R/2 \le r \le R$) respectively.  We define $\Lambda$ such that one has $\langle r^3 \rangle = \Lambda R^3$, with $\Lambda = 9 / 16$ for $M \le \Msun$ (outer shell convection) and $\Lambda = 1 / 16$ for $M > \Msun$ (core convection).  In Kolmogorov turbulence, the turbulent speed scales as $v(l) = (l \epsilon)^{1/3}$.  Putting everything together, we obtain $C(0)^{1/2} \propto \rho V_\mathrm{turb} R^{3} l^{-2} \epsilon$ and hence
\begin{equation}\label{eq:hRMS_scaling}
    \hrms = \frac{0.77 \, G \, \rho \, \Lambda R^6 \, m \, \epsilon}{c^{5} \, d \, l^2 \, M}~,
\end{equation}
where $M$ and $m = \rho V_\mathrm{turb}$ are the zero-age main-sequence and convective-zone masses respectively.

The power spectral density of the gravitational-wave strain, $S(f)$, is the Fourier transform of its autocorrelation function [see Appendix B in \citet{lasky13} for details],
\begin{equation}\label{eq:PSD_definition}
    S(f) = \int_{-\infty}^\infty d\tau \, \exp(i 2 \pi f \tau) C(\tau)~.
\end{equation}
Equation \eqref{eq:PSD_definition} holds when the decoherence time $\tau_c = 0.35 \etaks^{-1}$ is much shorter than the observation time $T_\mathrm{obs}$ \citep{melatos10, lasky13}.  Combining equations \eqref{eq:autocorrelation} and \eqref{eq:PSD_definition}, we obtain
\begin{eqnarray}\label{eq:PSD}
    \nonumber S(f) &=& \frac{5 \hrms^2 x^3}{8 \pi^3 f} \Bigg\{ \exp\left[ - 4 \pi x^2 \left(\frac{k_s}{k_d}\right)^{4/3} \right] \left[ 3 x^{-6} + 12 \pi x^{-4} \left(\frac{k_s}{k_d}\right)^{4/3} + 256 \pi^3 \left(\frac{k_s}{k_d}\right)^4 \right]\\
    &\;&\qquad\qquad\qquad - \exp\left(-4\pi x^2\right) \left( 3 x^{-6} + 12 \pi x^{-4} + 24 \pi^2 x^{-2} + 32 \pi^3 \right) \Bigg\}~.
\end{eqnarray}
where we define the rescaled frequency $x = f / \etaks$ for notational convenience.  The power spectral density peaks at $x \sim 1$, with $S(f)^{1/2} \propto f^2$ at low frequencies $x \lesssim 1$, $S(f)^{1/2} \propto f^{-2}$ at high frequencies $x \gtrsim 1$, and a sharp rollover at $f \gtrsim \eta(k_d)$.  Equation \eqref{eq:PSD} includes an additional factor $5/(4\pi)$ compared to equation \eqref{eq:autocorrelation}.  The latter expression applies purely to the $\ell = m = 2$ mode and an optimal (i.e. signal maximizing) orientation ($|T^{B2,22}_{jk}|^2 = 1$).  By contrast, equation \eqref{eq:PSD} contains the five modes $\ell = 2, |m| \le 2$, all of which have the same autocorrelation, and the tensor product is averaged over all possible sky locations and orientations of multiple sources, with
\begin{equation}
    \frac{1}{4\pi} \int_{-1}^{1} d(\cos\theta) \int_0^{2\pi} d\phi \; \left\langle T_{jk}^{B2,2m} T_{jk}^{B2,2m\star} \right\rangle = \frac{1}{4\pi}
\end{equation}
for fixed $m$, summing over $j$ and $k$, where $\theta$ and $\phi$ are the latitude and longitude of the observer relative to the source.

\subsection{Convective power}\label{sec:convective_power}
We assume for simplicity that the stellar luminosity $L$ is transported mechanically within the convective zone of a main-sequence star.  Hence energy is injected into the turbulence at a normalized rate $\epsilon = L/m$.  Although $m$, the mass enclosed within the convective zone, is a function of $M$, with $0 \le m(M) \le M$ [e.g., see Figure 22.7 in \citet{kippenhahn90}], we assume a uniform value $m = 0.3 M$ for simplicity.  The factor $m^{-1}$ in $\epsilon$ cancels with the factor $\int dV \rho = m$ in equation \eqref{eq:hRMS_scaling}, so that the wave strain $\hrms \propto m R \epsilon / d \propto L R / d$ behaves well for all $m$.  Stars with mass $M \gtrsim \Msun$ have a convective core and radiative outer shell, while stars with $M \lesssim \Msun$ contain a radiative core and convective outer shell \citep{kippenhahn90}.  Most of the luminosity (50\% for a $1 \Msun$ star and 90\% for a $10 \Msun$ star) is generated in the inner 10\% of the star by mass \citep{kippenhahn90}, so we approximate $L$ as constant and equal to its photospheric value throughout the convective zone.  A conservative reader may choose to reduce $L$ and hence $\hrms \propto \epsilon \propto L$ modestly to allow for this approximation; doing so does not significantly affect any of our conclusions.

The stirring and viscous dissipation scales of the turbulence are
\begin{equation}
    \label{eq:k_s} k_s = \frac{2 \pi}{l}~,
\end{equation}
and
\begin{equation} \label{eq:k_d}
    k_d = \left( \frac{8 \epsilon}{27 \nu^3} \right)^{1/4} \gg k_s,
\end{equation}
where $\nu$ is the kinematic viscosity.  We adopt $\mathrm{Re} = v H / \nu \sim 10^{10}$ as a typical Reynolds number, where $v = (l \epsilon)^{1/3}$ is the typical turbulent flow speed and $H$ is the hydrostatic scale length,
\begin{equation}
    H = c_s^2 / g~,
\end{equation}
with $g = G M / R^2$ and $c_s^2 = k_B T / \mu$, where $k_B$ is Boltzmann's constant, $\mu$ is the mean molecular mass, taken to equal the proton mass, and $T$ is the temperature in the convection zone.  In mixing length theory \citep[e.g.,][]{kippenhahn90}, $l$ is a free parameter, usually represented as,
\begin{equation}
    l = \alpha_\mathrm{MLT} H~,
\end{equation}
where $\alpha_\mathrm{MLT}$ is a constant, which can be determined from observations or simulations.  Table 4 in \citet{arnett09}, assembled from simulation data, implies $1.5 \lesssim \alpha_\mathrm{MLT} \lesssim 4$.  We take $\alpha_\mathrm{MLT} = 2$ throughout this paper and require the largest eddies to fit inside the star, viz. $l = \mathrm{min}(\alpha_\mathrm{MLT} H, R)$.

\subsection{Stellar mass-radius-luminosity relations}\label{sec:stellar_mass_radius_luminosity_relations}

The idealized model in Section \ref{sec:power_spectral_density} reduces, through $\epsilon$ and $\rho$, to a one-parameter function of the zero-age main-sequence mass $M$.  The radius and luminosity are related to $M$ through standard piecewise power-law fits to observations \citep{kippenhahn90, salaris06}:
\begin{equation}
    \label{eq:mass-radius} \frac{R(M)}{\Rsun} = \left(\frac{M}{\Msun}\right)^\alpha~,
\end{equation}
and
\begin{equation}
    \label{eq:mass-luminosity} \frac{L(M)}{L_\odot} \propto \left(\frac{M}{\Msun}\right)^\beta~,
\end{equation}
with
\begin{eqnarray}
     \alpha &=& \left\{ \begin{array}{ll} 0.80~, &\quad M / \Msun < 1~, \\ 0.57~, &\quad M / \Msun > 1~,\end{array} \right. \\
     \label{eq:mass-luminosity_beta} \beta &=& \left\{ \begin{array}{ll} 2.6~, &\quad M / \Msun < 0.5~, \\ 4.5~, &\quad 0.5 < M / \Msun < 2~, \\ 3.6~, &\quad 2 < M / \Msun < 20~, \\ 1.0~, &\quad M / \Msun > 20~.\end{array} \right.
\end{eqnarray}
The solar values are $\Msun = 2.0 \times 10^{30}$ kg, $\Rsun = 7.0 \times 10^{8}$ m, and $L_\odot = 3.8 \times 10^{26}$ W.  The average temperature in the convection zone approximately satisfies $T \propto M / R$ from virial equilibrium \citep[e.g.,][]{kippenhahn90}, i.e.,
\begin{equation}
    \label{eq:temperature} T(M) = T_0 \frac{M}{\Msun} \left[\frac{R(M)}{\Rsun}\right]^{-1}~,
\end{equation}
with
\begin{equation}
    \label{eq:T0}
    T_0 = \left\{ \begin{array}{ll} 2 \times 10^6 \textrm{ K}~, &\quad M / \Msun \le 1~, \\ 1 \times 10^7 \textrm{ K}~, &\quad M / \Msun > 1~.\end{array} \right.
\end{equation}
The jump in $T_0$ at $M = \Msun$ is physical; it reflects the sharp transition from outer shell to core convection seen in Figure 22.7 in \citet{kippenhahn90}.  The two values for $T_0$ in equation \eqref{eq:T0} refer to the base of the outer convection zone ($r \approx 0.7 \Rsun$; $M \le \Msun$) and the core ($r = 0$; $M > \Msun$).

\subsection{Mass quadrupole}\label{sec:mass_quadrupole}

In Section \ref{sec:power_spectral_density}, we calculate $S(f)$ assuming the current quadrupole dominates.  In neutron stars, where the density perturbations are small but the turbulent flow speed can reach a significant fraction of the speed of light, this is a good assumption \citep{melatos10}.  For stellar convection, the typical flow speed is lower (e.g., $v \sim 10^2$ m s$^{-1}$ in the Sun), and the mass quadrupole assumes heightened importance.

We estimate the ratio of the mass and current quadrupole wave strains as follows.  One has $|\Itwom| \sim I \delta\rho / \rho$ for subsonic density perturbations $\delta\rho \sim \rho v^2 / c_s^2$, where \Itwom is the mass quadrupole moment, $I$ is the unperturbed moment of inertia, and $c_s$ is the sound speed.  One also has $|\Stwom| \sim I v$ from equation (10) in \citet{melatos10}.  Both estimates contain a numerical pre-factor which arises from the \Ylm-weighted average over turbulent cells, which decreases with the number of cells but is similar for both \Itwom and \Stwom.  Current quadrupole terms in the wave strain expansion \eqref{eq:hTT} have an additional factor of $c^{-1}$ relative to the equivalent mass quadrupole terms, implying that the power spectral densities emitted by the mass and current quadrupoles are in the ratio
\begin{equation}\label{eq:mass_quadrupole_correction}
    \frac{S_\mathrm{mass}(f)}{S(f)} \approx \frac{|\Itwom|^2}{|\Stwom|^2/c^2} \approx \left( \frac{v c}{c_s^2} \right)^2~.
\end{equation}
One finds $S_\mathrm{mass}(f) / S(f) \lesssim 2$ for all $M$ and $S_\mathrm{mass}(f) > S(f)$ only for $10 \lesssim M / \Msun \lesssim 100$.  Hence the mass quadrupole boosts the overall gravitational-wave strain somewhat but does not make a significant difference to the final result.

An exact calculation of the mass-quadrupole contribution to $h_\mathrm{rms}$ is possible in principle within the framework set down by \citet{melatos10}, but it is not easy.  To appreciate why, recall that \Slm is proportional to the $Y^{\ell m}$-weighted volume integral of $\mathrm{curl}\,\vv$, for an incompressible fluid, so $C(\tau)$ is quadratic in $\vv$, i.e., it is proportional to a second-order unequal-time correlator of the form
\begin{equation}
    \langle v_i(\kk, t) v_j(\kk', t)^* \rangle~,
\end{equation}
where $\kk$ is the Fourier wavenumber.  In contrast, \Ilm is proportional to the volume integral of the instantaneous density perturbation $\delta\rho$.  For nearly incompressible flow, the secular terms in the Navier-Stokes equation give
\begin{equation}
    \delta\rho \propto v_i(\kk, t) v_j(\kk', t)^*~.
\end{equation}
Hence $C(\tau)$ is quartic in $\vv$, i.e., it is proportional to a fourth-order unequal-time correlator of the form
\begin{equation}
    \langle v_i(\kk, t) v_j(\kk', t)^* v_k(\kk'', t) v_l(\kk''', t)^* \rangle~.
\end{equation}
Fourth-order correlation functions are imperfectly known in Kolmogorov turbulence in a standard, Navier-Stokes fluid \citep{comte-bellot71, dong08}, depending sensitively on the boundary conditions and the nature of the driver.  They are even less understood in the context of viscous convection and lie outside the scope of this paper.

\subsection{Near zone}\label{sec:near_zone}

The Earth lies inside the Sun's gravitational-wave near zone for wavelengths $\lambda$ satisfying $\lambda / (2 \pi \dsun) = (f / 3 \times 10^{-4}~\mathrm{Hz})^{-1} \gg 1$, where \dsun is the Earth-Sun distance.  As the Sun is a convective star, it contributes strongly to the stochastic signal analyzed in this paper.

Inside the near zone, quadrupolar metric perturbations scale more steeply with distance ($\propto d^{-5}$) than in the wave zone ($\propto d^{-1}$) and are therefore easier to detect.  The multipole formula \eqref{eq:hTT}, which applies for $d \gtrsim \lambda / 2\pi$, does not capture this behavior.  To estimate the near-zone enhancement, we follow \citet{cutler96} and \citet{polnarev09}, who calculated the response of an interferometer to the gravitational perturbations created by Solar oscillations.  The metric in the vicinity of the Sun in the weak-field limit can be written \citep{polnarev09}
\begin{eqnarray}
    ds^2 &=& (\eta_{\mu \nu} + h_{\mu \nu}) dx^\mu dx^\nu \\
    \label{eq:Polnarev_metric} &=& \left(1 + \frac{2 U}{c^2}\right) c^2 dt^2 - \left(1 - \frac{2 U}{c^2}\right) dx_i dx^i + h_{ij}^\mathrm{GW} dx^i dx^j~,
\end{eqnarray}
where the perturbations $|h_{\mu \nu}| \ll 1$ to the Minkowski metric $\eta_{\mu \nu} = \mathrm{diag}(1, -1, -1, -1)$ are small, $U$ is the Newtonian gravitational potential, $h_{ij}^\mathrm{GW}$ is the radiative perturbation [i.e. the part which transports energy radially outwards, as in equation \eqref{eq:hTT}], and Greek (Roman) indices run over space-time (space) coordinates.

In the near zone, equation \eqref{eq:Polnarev_metric} is dominated by its quasistatic Newtonian part,
\begin{equation}\label{eq:deltaU}
    h_{\mu \nu}^\mathrm{N} = - \frac{2 G}{c^2 d^3} \delta_{\mu \nu} \sum_{\ell = 2}^\infty \sum_{m=-2}^2 \Ilm(t) \Ylm(\theta, \phi)~.
\end{equation}
We focus on the leading $\ell=2$ term and write it in terms of $\Stwom$ using equation \eqref{eq:mass_quadrupole_correction}, viz. $\Itwom \approx v \Stwom / c_s^2$.  In comparison, for stellar oscillations, one has $\Itwom(t) = \Msun \Rsun^2 \sum_n J_{nm}$, where $J_{nm} \propto e^{i \omega_n t}$ are dimensionless mass quadrupole moments for oscillation (e.g. g- and p-) modes with frequency $\omega_n$ \citep{cutler96,polnarev09}.  

The arm-length change $\delta L$ between two arms directed along the unit vectors $\nn$ and $\mm$ for a LISA-like interferometer with baseline $L$ is given by \citep{cutler96}
\begin{equation} \label{eq:h_nabla_nabla_deltaU}
    \delta L \propto \int_0^t dt' \int_0^{t'} dt'' \; (n^a n^b + m^a m^b) \nabla_a \nabla_b \delta U~,
\end{equation}
where $\delta U$ is the time dependent part of $U$, which is related to the wave strain in equation \eqref{eq:deltaU} by $h_{\mu \nu}^\mathrm{N} = 2 \delta U \delta_{\mu\nu} / c^2$ \citep{polnarev09}.  The wave strain $\propto \delta L$ in equation \eqref{eq:h_nabla_nabla_deltaU} scales as $d^{-2} \delta U \propto d^{-5}$ multiplied by a complicated angular dependence.  Hence the strain arising from $h_{\mu \nu}^\mathrm{N}$ rises more steeply with decreasing distance than the far-zone strain ($\propto d^{-1}$) given by equation \eqref{eq:hTT}, if the latter is extrapolated naively into the near zone.  Even when the radiative perturbation $h_{ij}^\mathrm{GW}$ is extrapolated correctly into the near zone, by one less power of $d$ than $h_{\mu \nu}^\mathrm{N}$ (as for an electromagnetic antenna), it remains smaller than $h_{\mu \nu}^\mathrm{N}$ by a factor $2 \pi d / \lambda$, becoming comparable at the boundary between the near and far zones.  The near-zone-corrected power spectral density is dominated by $h_{\mu \nu}^\mathrm{N}$ and satisfies
\begin{equation} \label{eq:S_near}
    S_\mathrm{near}(f) \approx \left( \frac{2 \pi \, d \, f}{c} \right)^{-8} \; S_\mathrm{mass}(f)~.
\end{equation}
Equation \eqref{eq:S_near} contains the same enhancement factor identified in equations (20) in \citet{polnarev09}.  To calculate the numerical constant in front of the factor $(2 \pi d f / c)^{-8}$, whose exact value depends in part on how one averages over detector orientation and source location for the observational strategy in question, we refer the reader to \citet{polnarev09}.  Note that equation \eqref{eq:S_near} only applies within the near zone; for $f > c / 2 \pi d$, the uncorrected $S(f)$ given by equation \eqref{eq:PSD} should be used.

Equation \eqref{eq:Polnarev_metric} neglects vector peturbations which may be present due to vorticity.  The metric is simplified to terms which dominate in the near and far zones.  Scalar terms dominate in the near zone, where they scale most steeply with distance from the source.  Tensor terms (gravitational radiation) dominate in the far zone \citep{polnarev09}.  Vector terms may contribute comparably to the scalar and tensor terms in the intermediate zone (distance from source $~$ wavelength) and change our results by a factor of order unity.  This effect is insignificant for the universal background but is potentially important for the Sun, where the Earth lies in the intermediate zone for frequency $f \sim 3 \times 10^{-4}$~Hz and should be included in refined calculations in future, if the prospects for detection improve.

\subsection{Detection threshold} \label{sec:detection_Sf_threshold}
In this section, we derive the threshold power spectral density required for detection.  A cross-correlation search, the method of choice for a stochastic background, relies upon the assumption that instrumental noise is uncorrelated between multiple antennas, while the stochastic signal is correlated.  As the observation time increases, the average noise decreases relative to the signal; in principle, detection is guaranteed after a sufficiently long observation.

For an isotropic signal and stationary, Gaussian noise, the signal-to-noise ratio is expressed as \citep{allen99}
\begin{equation} \label{eq:SNR_OmegaGW}
    \sigma \approx \frac{3 \, H_0^2 \, T_\mathrm{obs}^{1/2}}{10 \pi^2} \left[ \int_{-\infty}^\infty df \; \frac{\gamma^2(f) \, \Ogwsquared(f)}{f^6 \, S_1(f) \, S_2(f)} \right]^{1/2}~,
\end{equation}
where $H_0$ is Hubble's constant, \Tobs is the observation time, $\gamma(f)$ is the detector overlap function, $S_1(f)$ and $S_2(f)$ are the noise power spectral densities of the two detectors, and $\Ogw(f)$ represents the gravitational-wave energy density as a fraction of the closure energy density of the Universe per logarithmic frequency.  For an individual source like the Sun, the assumption of isotropy does not hold exactly, but it may hold approximately if a LISA-like interferometer achieves ergodic coverage of the sky over a long enough observation.  Relaxing the isotropy assumption falls outside the scope of this work.  By relating $\Ogw(f)$ to the power spectral density, according to \citep{sathyaprakash09}
\begin{equation} \label{eq:OmegaGW_2_PSD}
    S(f) = \frac{3 H_0^2 \Ogw(f)}{10 \pi^2 f^3}~,
\end{equation}
one obtains
\begin{equation} \label{eq:SNR_PSD}
    \sigma^2 = 2 \Tobs \int_0^\infty df \; \frac{\gamma^2(f) S^2(f)}{S_1(f)S_2(f)}~.
\end{equation}
A handy way to visualize whether the predicted background $S(f)$ in equation \eqref{eq:PSD} is detectable without performing the integral in equation \eqref{eq:SNR_PSD} is to define an effective, frequency-dependent signal-to-noise ratio $\sigma_\mathrm{eff}(f)$ per log frequency implicitly via
\begin{equation}
    \sigma^2 = \int_0^\infty d(\log{f}) \; \sigma_\mathrm{eff}^2(f)~,
\end{equation}
with
\begin{equation} \label{eq:rho_eff}
    \sigma_\mathrm{eff}^2(f) = \frac{2 f \Tobs \gamma^2(f) S^2(f)}{S_1(f)S_2(f)}~.
\end{equation}
In a full data analysis exercise, a detection requires that the integrated signal-to-noise ratio, given by equation \eqref{eq:SNR_PSD}, exceeds a specified threshold, e.g. $\sigma > \sigma_\mathrm{th}$.  Roughly speaking, this occurs when $\sigma_\mathrm{eff}(f)$ exceeds $\sigma_\mathrm{th}$ over approximately one decade in $f$, centered on the frequency where $\sigma_\mathrm{eff}(f)$ peaks.  For two colocated detectors with uncorrelated detector noise and identical spectral noise density $S_h(f)$, equation \eqref{eq:rho_eff} with $\sigma_\mathrm{eff} > \sigma_\mathrm{th}$ translates into the rule-of-thumb detectability condition
\begin{equation} \label{eq:approx_threshold}
    S(f) > \sigma_\mathrm{th} \left( 2 \, f \, \Tobs \right)^{-1/2} \; S_h(f)~.
\end{equation}
This inequality is equivalent to equation (136) in \citet{sathyaprakash09}, for $S(f)$ instead of \Ogw.

\section{Individual sources}\label{sec:individual_sources}

In this section, we calculate the gravitational-wave spectrum radiated by stars of different masses.  Upon combining equations \eqref{eq:eddy_turnover}, \eqref{eq:hRMS_scaling}, \eqref{eq:PSD}, and \eqref{eq:k_s}--\eqref{eq:T0}, the power spectral density $S(f)$ depends only on $M$ and $d$.

\subsection{Representative examples}\label{sec:representative_examples}

\begin{figure}
    \plotone{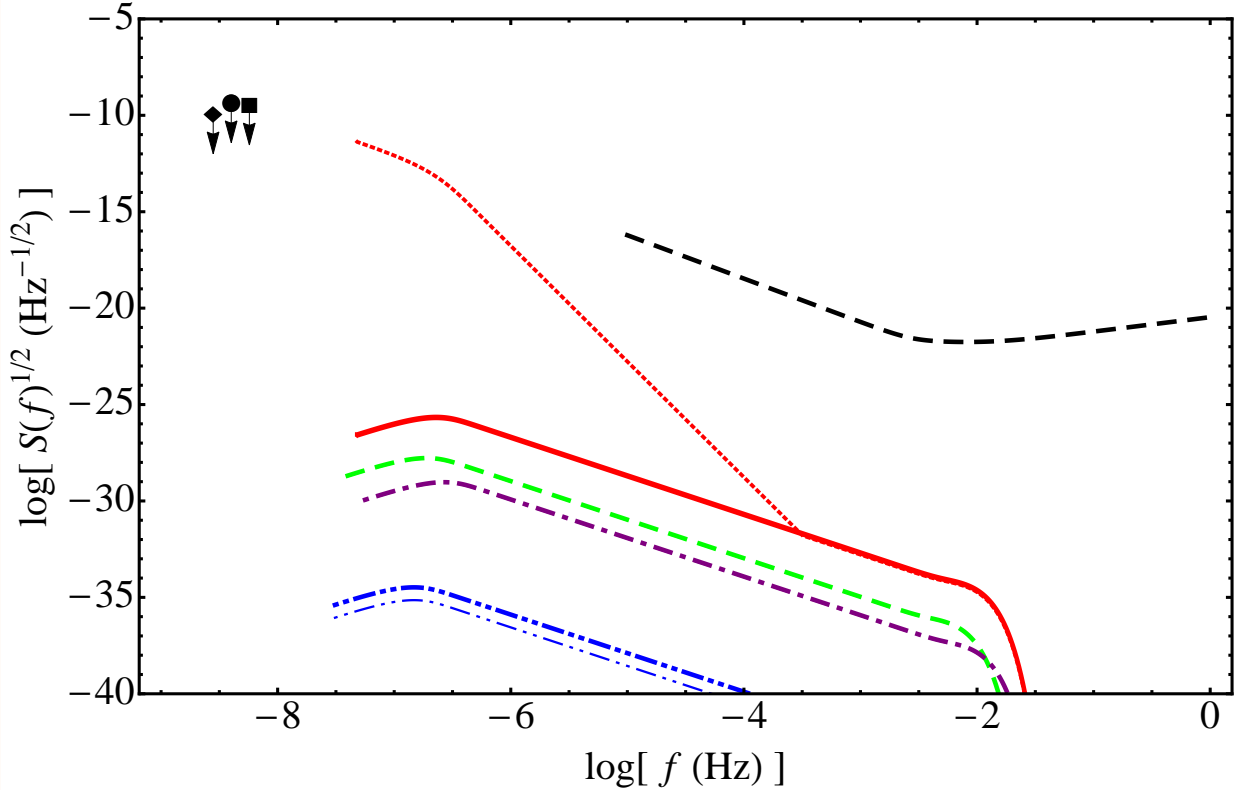}
    \caption{Gravitational-wave power spectral density $S(f)$ from stellar convection [plotted as $S(f)^{1/2}$] versus wave frequency $f$ for a selection of representative sources (mass, distance): the Sun ($\Msun$, $\dsun$) (solid red), the larger star in $\eta$-Carinae (120 $\Msun$, 2 kpc) (dashed green), a 0.25 $\Msun$ star at 10 pc (dash-double-dotted blue), and a 10 $\Msun$ star at 100 pc (dash-dotted purple).  Three components of the signal are plotted: the current quadrupole $S(f)^{1/2}$ from equation \eqref{eq:PSD} (thick curves), the estimated mass quadrupole $S_\mathrm{mass}(f)^{1/2}$ from Section \ref{sec:mass_quadrupole} (thin curves), and the near-zone correction for the Sun, $S_\mathrm{near}(f)^{1/2}$ (dotted red curve).  For comparison we also show the threshold for detection with LISA (dashed black) assuming the noise curve from \citet{sathyaprakash09} as well as pulsar timing array upper limits: circle \citep{vanhaasteren11}, square \citep{demorest13}, and diamond \citep{shannon13}.\label{fig:single_sources}}
\end{figure}

Figure \ref{fig:single_sources} displays $S(f)^{1/2}$ for a number of real and hypothetical sources.  The predicted signals are compared to the detection threshold given by equation \eqref{eq:approx_threshold} for LISA (dashed black curve) \citep{sathyaprakash09} and for three independent upper limits from pulsar timing array data: circle \citep{vanhaasteren11}, square \citep{demorest13}, and diamond \citep{shannon13}.  The sources presented are the Sun ($\Msun$, $\dsun$) (solid red), the larger star in $\eta$-Carinae (120 $\Msun$, 2 kpc) (dashed green), a 0.25 $\Msun$ star at 10 pc (dash-double-dotted blue), and a 10 $\Msun$ star at 100 pc (dash-dotted purple), chosen to represent typical sources in these mass and distance ranges.  For each source we plot $S(f)^{1/2}$ (solid curve) and $S_\mathrm{mass}(f)^{1/2}$ (dashed curve).  For the Sun we also plot $S_\mathrm{near}(f)^{1/2}$ (dotted red curve).

As can be seen in Figure \ref{fig:single_sources}, the spectrum resembles a piecewise power law, with $S(f)^{1/2} \propto f^2$ for $f \lesssim \etaks$ and $S(f)^{1/2} \propto f^{-2}$ for $f \gtrsim \etaks$.  It peaks at $f_\mathrm{peak} = 0.48 \etaks$.  However, none of the sources displayed in Figure \ref{fig:single_sources} are close to the LISA threshold.  The Sun is the strongest source, even without the large near-zone correction, with $f_\mathrm{peak} = 2.3 \times 10^{-7}$~Hz and $S(f_\mathrm{peak})^{1/2} = 2.3 \times 10^{-26}~\mathrm{Hz}^{-1/2}$.  $\eta$-Carinae, thought to contain one of the most massive known stars ($M \sim 120 \Msun$) \citep{damineli96}, is the next best candidate for detection.  With the exception of the Sun, high mass stars produce the largest amplitude signals and, despite their scarcity and relatively greater distance from Earth, are the best candidate for detection.  They have shorter lives than lower mass stars but emit significantly more gravitational radiation over their lifetimes than lower mass stars, which live many times longer.

The near-zone power spectral density $S_\mathrm{near}(f)^{1/2}$ increases as $f^{-2}$ as $f$ decreases below $3 \times 10^{-4}~\mathrm{Hz}$.  Pulsar timing arrays are sensitive to a stochastic gravitational wave background at frequencies $\approx (10~\mathrm{yr})^{-1}$.  To date, no detection has been achieved but a number of upper limits have been published.  Upper limits are typically evaluated for a single frequency in each independent study and are often presented in terms of $(H_0 / 100~\mathrm{km s}^{-1}~\mathrm{Mpc}^{-1})^2 \Ogw(f)$ at that frequency.  We use equation \eqref{eq:OmegaGW_2_PSD}, and take $H_0 = 73~\mathrm{km}~\mathrm{s}^{-1}~\mathrm{Mpc}^{-1}$, to convert the pulsar timing array upper limits to power spectral density for comparison with $S(f)$.  Note that such a comparison is not exact; pulsar timing array limits are calculated under different assumptions, e.g. an isotropic distribution rather than a single source.

The dotted curve in Figure \ref{fig:single_sources} extrapolates the near-zone spectrum below $f_\mathrm{peak}$.  We see that $S_\mathrm{near}(f)^{1/2}$ is on course to pass close to current pulsar timing array upper limits.  However, the spectrum flattens near the turnover frequency corresponding to the length scale $l = 2 \pi / k_s$, i.e. $f_\mathrm{peak} = 0.48 \eta(k_s) \approx 2.3 \times 10^{-7}~\mathrm{Hz}$ for the Sun; the largest and hence slowest eddies have $l \lesssim R$ and $f \gtrsim f_\mathrm{peak}$.  The assumptions in our model and hence the near-zone correction are suspect at frequencies beyond those represented physically in the Kolmogorov model, so we truncate the spectra in Figure \ref{fig:single_sources} at $f \le 0.1 \etaks$.

\subsection{Scalings}\label{sec:scalings}
We also explore briefly how $S(f)$ varies with the stellar parameters cited in Section \ref{sec:power_spectral_density}.  The spectrum resembles a piecewise power law, with $S(f)^{1/2} \propto f^2$ for $f \lesssim \etaks$ and $S(f)^{1/2} \propto f^{-2}$ for $f \gtrsim \etaks$, ignoring the near-zone correction.  It peaks at $f_\mathrm{peak} = 0.48 \etaks$, which for the Sun is $2.3 \times 10^{-7}~\mathrm{Hz}$.  Viscosity truncates the spectrum sharply at $f \gtrsim \eta(k_d)$, but this high-frequency rollover is unimportant for detection.

The root-mean-square wavestrain \hrms and decorrelation frequency \etaks scale with the variables describing stellar convection as $\hrms \propto L R^3 l^{-2} d^{-1}$ and $\etaks = \epsilon^{1/3} k_s^{2/3} \propto L^{1/3} m^{1/3} \, l^{2/3}$, if we substitute $\epsilon = L/m$ into equation \eqref{eq:hRMS_scaling}.  In Section \ref{sec:convective_power}, we assume $m \propto M$ and $l = \alpha_\mathrm{MLT} H \propto R$, whereupon the scalings simplify to $\hrms \propto L R d^{-1}$ and $\etaks \propto L^{1/3} M^{1/3} R^{2/3}$, with $R(M)$ and $L(M)$ taken from equations \eqref{eq:mass-radius}--\eqref{eq:mass-luminosity_beta}.  For the most massive stars ($M > 20 \Msun$), which radiate most strongly, we find $\hrms \propto M^{1.57}$ and $\etaks \propto M^{1.05}$.  For the least massive stars ($M < 0.5 \Msun$), which radiate weakly, we find $\hrms \propto M^{3.4}$ and $\etaks \propto M^{1.73}$.  In their three-dimensional simulations, \citet{arnett09} found that the maximum eddy size equals the thickness $\Delta R$ of the convection zone, therefore an alternative choice for the length scale of the largest eddies is $l = \Delta R$.  Increasing (decreasing) $l$ causes $\hrms \propto l^{-2}$ to decrease (increase) and $\etaks \propto l^{-2/3}$ to decrease (increase).

\section{Stochastic background}\label{sec:stochastic_background}
In this section, we calculate the stochastic gravitational-wave background produced by all the convective stars in the Universe.  Following general practice, we quantify the background in terms of its dimensionless energy density per logarithmic frequency interval \citep{sathyaprakash09},
\begin{eqnarray}
    \label{eq:OmegaGW_definition} \Ogw(f) &=& \frac{1}{\rho_c} \frac{d \rho_\mathrm{gw}}{d \ln f} \\
    \label{eq:OmegaGW_general} &=& \frac{1}{\rho_c c^2} \int_0^\infty \frac{dz \, n(z)}{1+z} \left( f_e \frac{dE}{df_e}\right)_{f_e = (1+z)f}~,
\end{eqnarray}
where $\rho_\mathrm{gw} c^2$ is the gravitational-wave energy density, $\rho_c c^2$ is the total energy density in a flat universe, $n(z)$ is the comoving number density of sources at redshift $z$, $f_e$ is the gravitational-wave frequency in the emitted frame, and $(dE/df_e)df_e$ is the gravitational-wave energy emitted by a source in the frequency interval $f_e$ to $f_e + df_e$.  We evaluate equation \eqref{eq:OmegaGW_general} for stars in the zero-age main-sequence mass range $(M, M+dM)$, so that $n(z)$ and $dE/df_e$ become functions of $M$, then integrate over $M$.
\begin{equation}\label{eq:OmegaGW_stellar}
    \Ogw(f) = \frac{1}{\rho_c c^2} \int dz \int dM \, \frac{n(z, M)}{1+z} \left( f_e \frac{dE(f_e, M)}{df_e}\right)\bigg|_{f_e = (1+z)f}~.
\end{equation}
Equation \eqref{eq:OmegaGW_general}, which applies to continuously emitting sources, has the same mathematical form as equation (5) in \citet{phinney01} for burst events (e.g., compact binary coalescences), but the physical interpretation of its factors is slightly different, as explained in Section II C of \citet{lasky13}.  For continuously emitting sources, $n(z,M)$ is the finite number of sources per unit mass, which each emit an infinitesimal amount of energy $dE/df dz$ during the redshift interval $(z, z+dz)$.  For burst sources, $n(z,M) dz$ is the infinitesimal number of impulsive events per unit mass occurring in $(z, z+dz)$, each emitting a parcel of finite energy $dE/df$.  In this paper, we choose the latter interpretation as a fair approximation because the main-sequence (and hence gravitational-wave-emitting) lifetime is much shorter than the lookback time to redshift $z$ for most stars.  The approximation breaks down for low mass stars with $M \lesssim \Msun$ at redshifts $z \gtrsim 1$, which are still alive and contributing to the background energy today.  However, the approximation is reasonable because these low mass stars produce only a tiny fraction of the overall signal; stars with $M < 10 \Msun$ contribute $\sim 10^{-4}$ of the total background.

We compute the stellar birth rate per unit redshift per unit mass $n(z, M) dz dM$ from the star formation rate $\dot{\rho}_*(z)$ and an initial mass function $\Phi(M)$,
\begin{equation}\label{eq:stellar_birth_rate}
    n(z, M) dz dM = \left[ \int_{M_\mathrm{min}}^{M_\mathrm{max}} dM M \Phi(M) \right]^{-1} \dot{\rho}_*(z) dz \Phi(M) dM~,
\end{equation}
where $M_\mathrm{min}$ and $M_\mathrm{max}$ are the smallest and largest main-sequence masses.  We adopt the Salpeter initial mass function, $\Phi(M) \propto M^{-2.35}$ with $M_\mathrm{min} = 0.1 \Msun$ and $M_\mathrm{max} = 125 \Msun$.  For the star formation rate $\dot{\rho}_*(z)$, we use the parametric fit to ultraviolet and far-infrared measurements out to $z\approx 6$ for a modified Salpeter IMF from \citet{hopkins06}.

The total gravitational wave energy per unit frequency emitted over a star's lifetime is given by \citep{lasky13}
\begin{equation}
    \label{eq:dEdf_definition} \frac{dE}{df} = \frac{\pi c^3 d_L^2 \tau_\mathrm{em}}{G} f^2 S(f)~,
\end{equation}
The emitting lifetime $\tau_\mathrm{em} = \min[\tau_\mathrm{life}(M), t_\mathrm{lb}(z)]$ is the minimum of the stellar nuclear lifetime $\tau_\mathrm{life}$ and the lookback time $t_\mathrm{lb}$, with $\tau_\mathrm{life} \propto M / L$ and ${\tau_\mathrm{life}}_\odot = 10^{10}$ yr.  The luminosity distance $d_L$ appears in equation \eqref{eq:dEdf_definition} but cancels a factor $d_L^{-2}$ in $S(f)$, so that the final result is independent of $d_L$.

\subsection{Detectability}\label{sec:detectability}

\begin{figure}
    \plotone{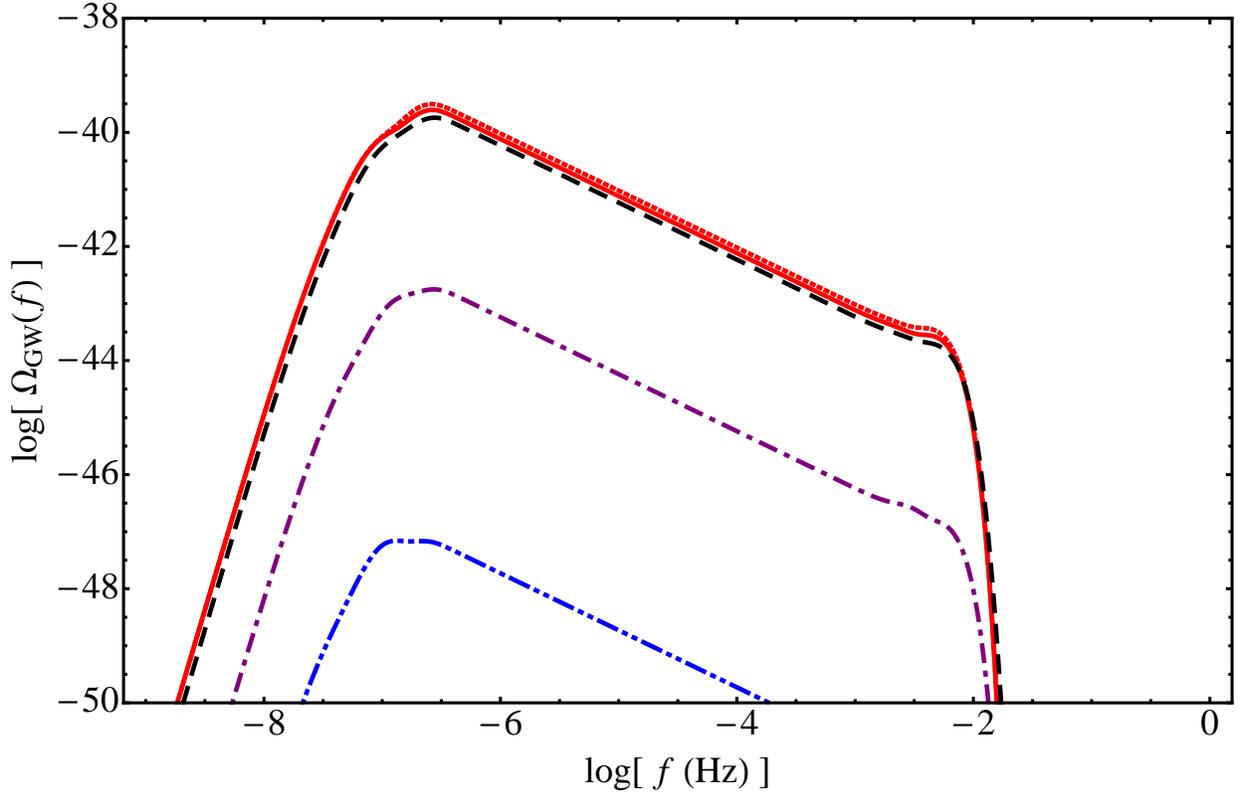}
    \caption{Total gravitational wave energy density per log frequency $\Ogw(f)$ (dimensionless) versus frequency $f$ (in units of Hz).  Solid and dotted curves (red in the online version) represent the current and mass quadrupole contributions respectively (see Section \ref{sec:mass_quadrupole}).  Dash-double-dotted (blue), dash-dotted (purple), and dashed (black) curves represent the energy density produced by stars in the mass ranges $0.1 \Msun \le M \le \Msun$, $\Msun \le M \le 10 \Msun$, and $10 \Msun \le M \le 100 \Msun$ respectively. \label{fig:universe}}
\end{figure}

Figure \ref{fig:universe} displays the energy density $\Ogw$ over the frequency range $10^{-9}~\mathrm{Hz} < f < 0.1~\mathrm{Hz}$ accessible by detectors from pulsar timing arrays to LISA.  The spectrum can be approximated as a piecewise power law, with $\Ogw(f) \propto f^7$ for $f \lesssim 10^{-7}~\mathrm{Hz}$ and $\Ogw(f) \propto f^{-1}$ for $10^{-7}~\mathrm{Hz}~\lesssim f \lesssim 0.01~\mathrm{Hz}$.  For $f \gtrsim 0.01~\mathrm{Hz}$, the spectrum cuts off due to viscosity.  The shape of the spectrum is the same as for neutron star turbulence \citep{lasky13}.

The stochastic background in Figure \ref{fig:universe} is too weak to be detected with current instruments.  At best, pulsar timing arrays and space-based interferometers are sensitive to backgrounds with $\Ogw \approx 1.3 \times 10^{-9}$ and $1.3 \times 10^{-14}$ at $f \approx 2.8$~nHz and $2.4$~mHz respectively.  To compare the isotropic stochastic background from multiple sources with the individual sources displayed in Figure \ref{fig:single_sources}, one can convert $\Ogw(f)$ to a power spectral density $S_\mathrm{gw}(f)$ using equation \eqref{eq:OmegaGW_2_PSD}.  Doing so reveals that $S(f)$ for both the Sun (with or without the near-field correction) and $\eta$-Carinae lie above the background, while the representative $10 \Msun$ and $0.25 \Msun$ stars fall below.  The former two objects give some idea of the largest local fluctuations from strong individual sources in the Milky Way above the mean background level generated by an isotropic stellar population.

\subsection{Comparison with other backgrounds}

\begin{figure}
    \plotone{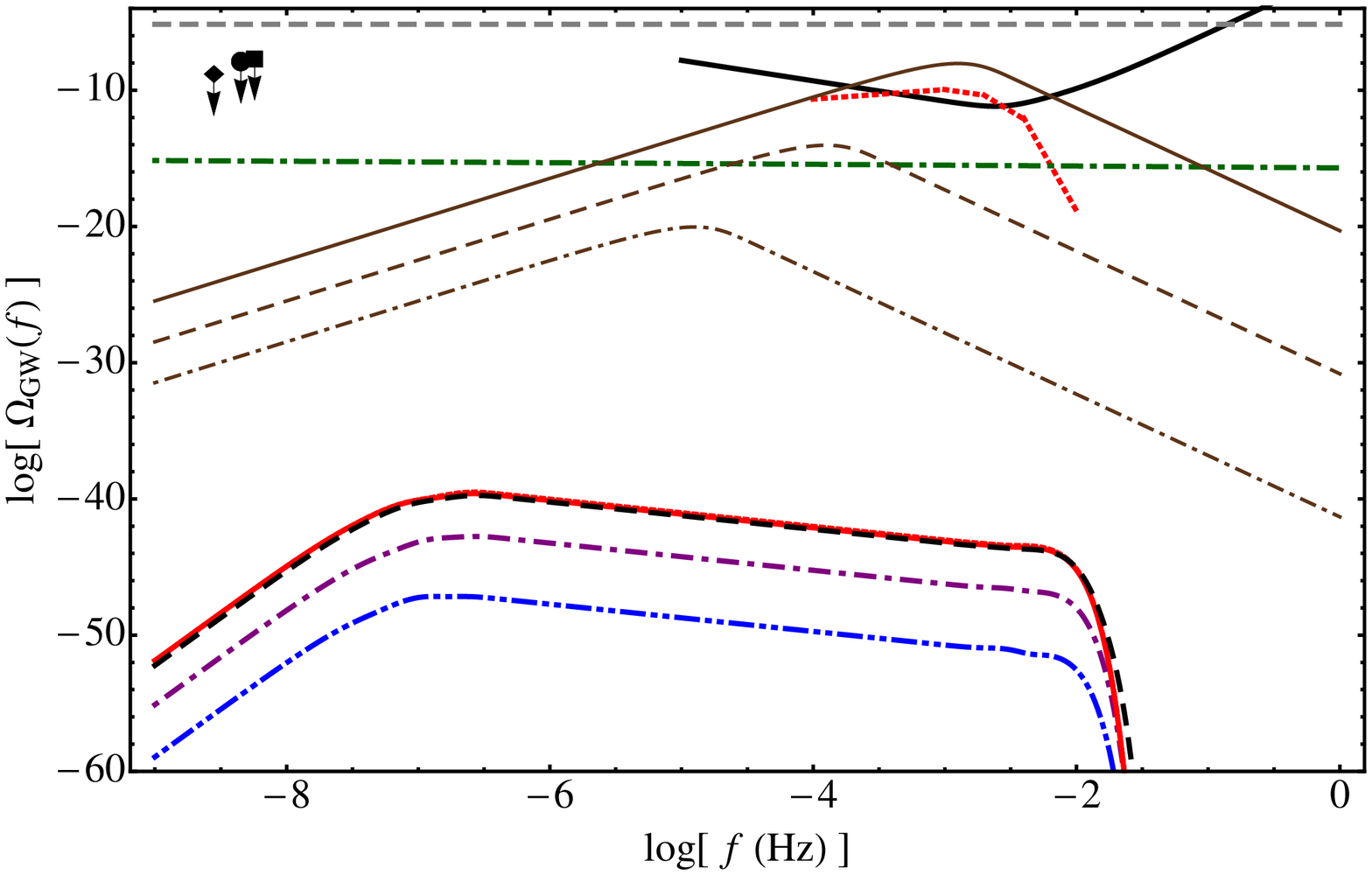}
    \caption{Total gravitational wave energy density per log frequency $\Ogw(f)$ (dimensionless) versus frequency $f$ (in units of Hz).  The figure reproduces the stellar convection curves from Figure \ref{fig:universe} and adds three extra stochastic sources for comparison.  The dotted curve (red in the online version) estimates the confusion noise from Galactic white dwarf binaries \citet{timpano06}.  The dash-dotted green curve represents the background from slow-roll inflation with $r = 0.2$ \citep{turner97}.  The thin brown curves represent the predicted relic gravitational wave background for primordial turbulence with Mach number $\mathcal{M} = 1$ (solid), $\mathcal{M} = 0.1$ (dashed), and $\mathcal{M} = 0.01$ (dash-dotted) from \citet{gogoberidze07}.  For comparison, we also show the LISA detection threshold (solid black) \citep{sathyaprakash09}, the pulsar timing array upper limits from Figure \ref{fig:single_sources} converted to $\Ogw$, and the LIGO upper limit on a frequency-independent background $\Ogw(f) < 6.9 \times 10^{-6}$ (dashed gray), set around 100 Hz \citep{abbott09}. \label{fig:universe_comparison}}
\end{figure}

Figure \ref{fig:universe_comparison} compares the predicted stellar convection spectrum against three other backgrounds which are often discussed:  confusion noise from Galactic white dwarf binaries \citep{timpano06}, relic gravitational waves from inflation \citep{turner97}, and relic gravitational waves from primordial turbulence \citep{gogoberidze07}.  For comparison, we also show the threshold for detection with LISA \citep{sathyaprakash09}, the upper limits from pulsar timing arrays, and the upper limit on a frequency-independent cosmological stochastic gravitational wave background, $\Ogw(f) < 6.9 \times 10^{-6}$, set by LIGO \citep{abbott09}.  It is important to note that the LIGO limit is set over the frequency band 41.5-169.25 Hz.  It is displayed here at a much lower frequency range purely as an interesting value for comparison.

Galactic white dwarf binaries are promising LISA sources, with thousands expected to be observable \citep{sathyaprakash09}.  At $f \lesssim$~mHz their signals are confusion limited; there are too many sources per frequency bin to resolve, creating a noise floor above the expected detector noise.  Figure \ref{fig:universe_comparison} displays an estimate of the confusion noise from Galactic white dwarf binaries, using the piecewise power-law fit in \citet{timpano06} for the 10\% background with the \citet{nelemans04} population model (red dotted curve).  Gravitational waves from inflation produce another background.  There are many choices of model \citep{starobinskii79, bar-kana94, turner97, smith06, easther07, barnaby12}.  Figure \ref{fig:universe_comparison} shows the background for a slow-roll inflation model (dash-dotted green curve) \citep{turner97}, with tensor-to-scalar ratio $r = 0.2$ corresponding to the best fit result from the BICEP2 experiment \citep{bicep14}.  Finally, we also plot in Figure \ref{fig:universe_comparison} the relic background from primordial turbulence \citep{kosowsky02, gogoberidze07}.  This signal depends on the total energy injected into the primordial plasma and the stirring scale of the turbulence, which can parameterized in terms of the Mach number $\mathcal{M} = (\epsilon / k_s)^{1/3} / c_s$.  Figure \ref{fig:universe_comparison} displays three versions of the primordial turbulence spectrum (thin brown curves) predicted by \citet{gogoberidze07} for $\mathcal{M} = 1$ (solid), $\mathcal{M} = 0.1$ (dashed), and $\mathcal{M} = 0.01$ (dash-dotted).  The $\mathcal{M} = 1$ spectrum rises above both the LISA threshold and the white dwarf confusion noise.

In Section \ref{sec:scalings}, we examine how the stellar convection spectrum scales with stellar parameters.  How do the results compare with neutron star turbulence \citep{melatos10,lasky13} and primordial turbulence \citep{kosowsky02,gogoberidze07}?  All three mechanisms assume isotropy and a Kolmogorov spectrum, where power is injected at rate $\epsilon$ and the largest (stirring) scale is $2\pi / k_s$.  The peak frequency and amplitude scale proportionally to $\epsilon^{1/3} k_s^{2/3}$ and $\epsilon^{1/2} k_s^{-1/2}$ respectively and can be related to different observables in each case.  Differences arise due to the means by which power is injected and the size of the largest turbulent eddies.  In stellar convection, energy is supplied by the star's luminosity, and the scale of the largest eddies is fitted from simulations but cannot exceed the physical size of the star.  Neutron star turbulence is similar, except that it is driven by an angular velocity shear $\Delta\Omega$ between the crust and core.  Both spectra scale as $\Omega_\mathrm{GW} \propto f^7$ and $\propto f^{-1}$ for frequencies below and above the peak frequency \citep{lasky13}, but neutron star turbulence peaks at a much higher frequency ($\sim$ 100 Hz) than stellar convection ($\sim \mu$Hz).  The peak frequency depends on the $\Delta\Omega$ distribution across the neutron star population and scales $\propto \Delta\Omega^7$ in an individual object.  The physical mechanisms which excite primordial turbulence, and the associated characteristic scales, are uncertain.  For example, an electroweak phase transition may provide sufficient energy to produce a stochastic background detectable by LISA \citep{apreda02,randall07,gogoberidze07}.  The spectrum also depends on the temperature and other properties of the early Universe at the time it is generated.  \citet{gogoberidze07} derived asymptotic limits for how the characteristic wave strain $h_c$ scales at frequencies above or below the peak frequency, finding $h_c \propto f^{1/2}$ and $h_c \propto f^{-13/4}$ respectively.  Converting from wave strain to $\Ogw$ for comparison with the scalings above, the equivalent results at low and high frequencies are $\Ogw \propto f^{3}$ and $\Ogw \propto f^{-9/2}$ respectively.

\section{Conclusion}\label{sec:conclusion}

Stellar convection and its associated, small-scale, Kolmogorov turbulence generates a stochastic gravitational wave signal.  The signal is guaranteed to exist and establishes an astrophysical noise floor below which other stochastic signals are undetectable.  Our calculations predict $S(f)^{1/2}$ for most individual Galactic sources to be $\gtrsim 14$ orders of magnitude below the LISA threshold.  The Sun is an exception.  Its spectrum peaks at $f_\mathrm{peak} = 2.3 \times 10^{-7}~\mathrm{Hz}$, where the far-zone power spectral density is $S(f_\mathrm{peak}) = 5.2 \times 10^{-52}~\mathrm{Hz}^{-1}$.  However, the Earth lies within the near zone of the Sun for frequencies $f < 3 \times 10^{-4}~\mathrm{Hz}$.  Metric perturbations scale more steeply with distance in the near zone ($\propto d^{-5}$) than in the far zone ($\propto d^{-1}$).  The near-zone power spectral density at the peak frequency is $S_\mathrm{near}(f_\mathrm{peak}) = 4.1 \times 10^{-27}~\mathrm{Hz}^{-1}$.  This falls in a gap in sensitivity between LISA and pulsar timing arrays.  Extrapolating the near-zone spectrum to lower frequency, we find that it is on course to rise above pulsar timing array upper limits.  However, we emphasize that the Kolmogorov model breaks down at $f \lesssim 0.1 \etaks \lesssim 10^{-8}~\mathrm{Hz}$, and our calculation of $S(f)$ assumes an isotropic background rather than a single source.  Any comparisons with pulsar timing array data in the future need to be considered in this light.  The Solar signal is a consideration for the design of future space-based interferometric detectors and pulsar timing array searches.

We make several simplifying assumptions when calculating $S(f)$ in Section \ref{sec:stochastic_gw_signal}.  The convective-zone mass is assumed to be $m = 0.3 M$.  Going from $m = 0.01 M$ to $m = M$ decreases the value of $f_\mathrm{peak}$ by a factor $\sim 5$ and $S(f_\mathrm{peak})$ by a factor $\sim 2 \times 10^3$.  The typical scale length of the largest eddies is taken from mixing length theory.  An alternative is to assume that $2\pi/k_s$ equals the depth of the convection zone.  If we take $l = 0.1 R$ instead of $l = \alpha_\mathrm{MLT} H$, the values of $f_\mathrm{peak}$ and $S(f_\mathrm{peak})$ are multiplied by factors of 1.4 and 0.83 respectively for the Sun and 4.6 and 0.46 for a $M = 100 \Msun$ star.  Going from $l = 0.01 R$ to $l = R$, the value of $f_\mathrm{peak}$ decreases by a factor $\sim 20$ and $S(f_\mathrm{peak})$ increases by a factor $\sim 20$.

The background energy density $\Ogw$ lies well below the sensitivity of LISA and pulsar timing arrays assuming the Salpeter initial mass function and \citet{hopkins06} star formation rate.  The Salpeter initial mass function overpredicts low mass stars, and does not evolve with redshift.  Population III stars formed in the early universe are thought to have masses $\gtrsim 100 \Msun$ \citep{abel02}.  A top-heavy initial mass function at high redshift produces more high-mass stars, and the inferred star formation rate required to predict the observed luminosity decreases \citep{hopkins06}.  As high mass stars generate the largest wave strain, a top-heavy initial mass function at high $z$ boosts the background.  More information on the absolute number and nuclear lifetimes of Population III stars is required to determine how significant their contribution might be.

\acknowledgements
We thank P. Lasky and V. Ravi for helpful discussions, and R. Sturani for feedback on the draft manuscript.  This research was supported by a Discovery Project grant from the Australian Research Council.


\end{document}